\documentclass[12pt, preprint]{aastex}

\usepackage{amsmath}
\usepackage{amssymb}

\newcommand{\emp}{\rm}

\newcommand{\dd}{\mathrm{d}}

\newcommand{\msun}{M_{\odot}}
\newcommand{\be}{\begin{equation}}
\newcommand{\ee}{\end{equation}}
\newcommand{\bea}{\begin{eqnarray}}
\newcommand{\eea}{\end{eqnarray}}

\shorttitle{Isolated halos}
\shortauthors{MacMillan, Henriksen, \& Widrow}

\begin{document}

\title{On Universal Halos and the Radial Orbit Instability}

\author{Joseph D. MacMillan, Lawrence M. Widrow, and Richard N. Henriksen} 
\affil{Department of Physics, Queen's University, Kingston, Ontario,
Canada, K7L 3N6} \email{macmilla@astro.queensu.ca,
widrow@astro.queensu.ca, henriksn@astro.queensu.ca}

\begin{abstract}
The radial orbit instability drives dark matter halos toward a
universal structure.  This conclusion, first noted by Huss, Jain, and
Steinmetz, is explored in detail through a series of numerical
experiments involving the collapse of an isolated halo into the
non-linear regime.  The role played by the radial orbit instability in
generating the density profile, shape, and orbit structure is
carefully analyzed and, in all cases, the instability leads to
universality independent of initial conditions.  New insights into the
underlying physics of the radial orbit instability are presented.

\end{abstract}



\section{INTRODUCTION}

Spherical infall models were introduced by \citet{gg72},
\citet{gott75}, and \citet{gunn77} to provide a simple theoretical
framework for understanding the formation of dark matter halos.  The
essence of these models is the radial collapse of successive spherical
shells onto a primordial density perturbation.  If the initial density
perturbation is scale-free the system evolves to a self-similar state
and may be followed into the non-linear regime with arbitrary accuracy
\citep{fg84, bert85}.

It is widely accepted that real dark halos form through hierarchical
clustering.  N-body methods provide a direct means to study this
rather complex process; the initial conditions are calculable from
well-motivated theories of the early Universe and the dynamical
equations involve only gravity and the assumption that dark matter is
collisionless.

Simulated halos exhibit a number of features that suggest universality
in structure and formation history.  For example, the density profiles
of simulated halos across many orders of magnitude in mass can be
characterized by a simple two-parameter fitting formula.  This result
was first discovered by \citet{nfw} who introduced what has come to be
known as the NFW profile:

\begin{equation}
\label{eq:nfw}
\rho(r) = \frac{\rho_s}{r/r_s\left (1 + r/r_s\right )^2}~.
\end{equation}
The concentration parameter $c$ associated with this profile is
$R_{\rm vir}/r_s$, where normally $R_{\rm vir}=R_{200}$, the subscript
indicating the mean overdensity relative to the background.

Similarly, \citet{bull01} showed that the cumulative angular momentum
distribution can be fit by a two-parameter formula to be discussed
below.  Moreover, \citet{taylor01} (see also \citet{dehnen05}),
found that $\rho/\sigma^3$, which they argue is a proxy for the
``phase-space density'', is well-approximated by a pure power-law in
radius.

The self-similar infall models exhibit some of the features of
hierarchically formed halos such as gradual growth in mass and size
and virialization from the inside out.  But do these models capture
the essential physics of hierarchical clustering?  The naive answer is
no.  Many predictions made by the self-similar models are inconsistent
with results from simulations.  In particular, density profiles for
the halos found in the radial self-similar models are
well-approximated by pure power-laws in radius with a logarithmic
slope close to $-2$ (but see \citet{hw99} and \citet{lu} for $r^{-3}$
in the outer regions).  By contrast, the profiles of simulated halos
are shallower than $r^{-2}$ in the inner regions and steeper than
$r^{-2}$ in the outer regions \citep{frenk88, dub91, nfw}.

The orbit structure also presents a problem for the naive self-similar
models.  The velocity distribution of dark matter in simulated halos
varies smoothly from isotropic to radially-biased as one moves out
from the center to the virial radius.  By fiat, the orbits in the
naive self-similar models are purely radial. More general
self-similarity that permits complex angular dependence and velocity
anisotropy has been studied in \citet{henrik04}.  There it is shown
that in lowest order coarse-graining, the radial dependence of these
solutions is not changed by the angular dependence of the underlying
system (see equation A-4 of that paper).  Thus the reason for the
disagreement with the simulated profiles must be sought in the finer
grained correction terms, but the physical nature of these terms
remains unclear \citep{henrik06}.

We recall that hierarchical clustering involves a complex series of
merger and accretion events which are inherently aspherical processes.
Even if some of the progenitors of a given halo develop steep cusps
through approximately spherical infall, one might well imagine these
cusps becoming shallower through merging and subsequent relaxation.  A
recent paper by \citet{dehnen05a} suggests difficulties with this
picture.  Based on a theorem for phase mixing, Dehnen argues that in
the merger of two sub-halos, the steepest cusp survives.  If by chance
some of the progenitors of a halo have cusps as steep as $r^{-2}$
(that is, if they form through radial collapse from approximately
scale-free initial conditions) the final halo will have a
correspondingly steep cusp. The corrections to the self-similar infall
referred to in the previous paragraph can mitigate this problem, but
how are they effected?

Here we explore whether the missing ingredient in this picture is the
radial orbit instability (ROI) \citep{hjs}. The ROI, as the name
suggests, occurs in systems initially biased towards radial orbits.
The effect of the instability is to drive the shape of the system
toward that of a prolate-triaxial bar, to flatten the inner density
cusp, and to isotropize the orbits in the inner region.  Indeed, the
final density profile of a system that has undergone the ROI is very
similar to the NFW profile \citep{hjs}.  The ROI is a collective
effect and, as with violent relaxation \citep{ldb}, redistributes the
energy and angular momentum of the particles.
The net effect is to remove much of the dependence on initial
conditions and hence drive the system toward a universal structure.

The first hints of universality in gravitational collapse were found
by \citet{van82} and \citet{may84} who considered the collapse of an
initially spherical region.  Although their simulations did not
include the effects of cosmological expansion and used only 5000
particles, they were able to demonstate that a wide range of initial
conditions leads to an apparently universal state provided that the
initial state is far from equilibrium.  The mechanism appears to be
violent relaxation \citep{ldb} and the system reaches a final state
well-described by $R^{1/4}$-law \citep{deV}.

The potential relevance of the ROI in cosmological structure formation
was first noted by \citet{aguilar90} who found that a sufficiently
cold (i.e., radial) initial state is bar-unstable.  \citet{cm95}
carried out a series of simulations designed to explore the ROI in a
cosmological setting.  The question raised was why, in the presence of
the ROI, are there spherical galaxies and (presumably) nearly
spherical halos?  As in earlier studies, \citet{cm95} followed the
evolution of an initially spherical region but added the effects of
the Hubble expansion and cosmological fluctuations.  They found that
the ROI is a transient effect; that systems can develop bar-like
structures early on, but that these structures become more spherical
as hierarchical clustering continues.  We argue that those early
stages where the ROI is important are nevertheless crucial in
establishing the $r^{-1}$ density profile seen in the simulations.

The direct link between the ROI instability and the universality of
dark matter halos was made by \citet{hjs} who carried out a series of
simulations of both isolated halos and of properly cosmological
volumes.  They note that the NFW profile appears to be a universal
feature of all their simulated halos and that in the case of spherical
collapse, the profile is generated by the ROI.

In this paper, we investigate the gap between the self-similar
infall models and the cosmological simulations.  To this end, we
follow the collapse of the single, isolated density peak from the
quasi-linear phase through to the formation of a galaxy-sized halo.
The model for initial conditions is that of a smooth spherical
density perturbation with initial profile predicted by linear
perturbation theory \citep{bbks}.  A series of experiments is designed
to isolate, in turn; the effects of the ROI, of small-scale
perturbations, and of tidal fields.  We explore the connection between
the ROI and halo density profiles in  detail and consider other
indications of universality such as the angular momentum profile and
the phase-space density profile.

The outline of the paper is as follows: In Section 2, we review
essential results from the self-similar infall models, Dehnen's
``merging theorem'' for cuspy halos, and the ROI.  In Section 3 we
describe the initial conditions and numerical methods used in our
experiments and provide an overview of our series of simulations.  We
present a detailed analysis of the simulations in Section 4 and devote
Section 5 to a discussion of the ROI.  Our conclusions are presented
in Section 6.

\section{PRELIMINARIES}
\subsection{Self-similar infall models}

Early attempts to understand cosmological structure formation were
based on the spherical infall model \citep{gg72, gott75, gunn77} in
which matter from the cosmic background accretes slowly onto a
smooth, spherically symmetric density perturbation.  If the density
profile of the initial perturbation is scale-free, that is, if
$\delta\rho_i \propto r^{-\epsilon}$ ($\epsilon\in\{0,\,3\}$) the
system evolves to a self-similar state in the sense that its
phase-space distribution function takes the form

\begin{equation}
f\left (r,\,v_r,\,t\right ) = {\cal G}(t)
{\cal F}\left (r/r_*(t),\,v_r/v_*(t)\right )
\end{equation}

\noindent where $r_*$, $v_*$, and ${\cal G}$ are power-law functions
of time \citep{fg84, bert85}.  To determine the scaling relations,
note that at each time $t$, there is a particular mass shell that has
reached its maximum or turnaround radius and is beginning its collapse
towards the center.  The turnaround radius scales with time as $r_{\rm
ta}\propto t^\delta$ where $\delta = 2/3 + 2/\left (3\epsilon\right
)$.  The self-similar solutions of \citet{fg84} and \citet{bert85} are
derived by setting $r_* = r_{\rm ta}$ and $v_* = dr_{\rm ta}/dt$.  (See
\citet{hw99} for a discussion of the self-similar infall model based
on the collisionless Boltzmann equation.  This formulation allows for
more general types of self-similarity.)

The density profile in the self-similar state is well-approximated by
a power-law, $\rho(r)\propto r^{-\mu}$, where

\begin{equation}
\mu = \left\{\begin{array}{ll}
2/\delta & \mbox{if $\delta < 1$} \\
2 & \mbox{otherwise}~.\end{array}\right.
\end{equation}

\noindent In a cold dark matter dominated Universe, the primordial
density field is characterized by gaussian random perturbations with a
power-spectrum $P(k)$ and correlation function

\begin{equation}
\xi(r) = \frac{1}{\left (2\pi\right )^3}\int
P(k)e^{-i{\bf k}\cdot {\bf r}}d^3k~.
\end{equation}

\noindent Consider a $\nu\sigma$ peak of this field where $\sigma\equiv
\xi(0)^{1/2}$.  \citet{peebles84} argued that the mean density a
distance $r$ from the peak is given by

\begin{equation}
\label{peak}
\left \langle \frac{\delta \rho}{\rho}\right \rangle
= \frac{\nu\xi(r)}{\sigma}~.
\end{equation}

\noindent For a scale-free power-spectrum, $P(k)\propto k^{n}$,
$\xi(r)\propto r^{-\left (n+3\right )}$ and therefore $\mu = \left
(3n+9\right )/\left (n+4\right )$ for $n>-1$ and $\mu = 2$ for $n\le -1$
\citep{hs85}.

Eq.\,{\ref{peak} is valid for a $\nu\sigma$ {\it extremum} of the
primordial density field.  If one adds the additional constraint that
the region of interest is a {\it maximum} the density profile is
modified \citep{bbks}.  The implications of this modification in the
context of the spherical infall model were explored by \citet{rg87}.

Naive self-similarity can be extended to incorporate non-radial
motions.  \citet{sikivie97}, \citet{nusser01}, \citet{ledelliou03},
and \citet{asc04} include angular momentum into the infall model while
preserving self-similarity and spherical symmetry.  These conditions
require that the distribution of particle orbits is isotropic in the
plane tangent to the radial direction and that the specific rms
angular momentum at turnaround is proportional to $\sqrt{M_{\rm
ta}r_{\rm ta}}$.  Not surprisingly, angular momentum tends to soften
the inner cusp yielding a profile that shares some of the
characteristics of the NFW profile.  (See also \citet{henrik04} who
argues that flattening of a central density cusp to a core is a
natural final state of relaxation processes.)

Along rather different lines, \citet{ryden93} found self-similar
solutions for the collapse of an axisymmetric perturbation in an
expanding Universe.  These models can be prolate or oblate and appear
to have a density profile somewhat shallower than $r^{-2}$ in the
inner regions.

\subsection{Merging theorem for cuspy halos}

Recently, \citet{dehnen05a} argued that in the merger of two cuspy
halos, the steepest cusp always survives.  That is, the cusp of the
remnant halo has the same logarithmic slope as the steepest cusp
of the two progenitors.  The theorem is based on the observation
that the excess-mass function

\begin{equation}
D\left (f\right ) \equiv 
\int_{\overline{F}\left ({\bf x},\,{\bf v}\right ) > f}
\left [\overline{F}\left ({\bf x},\,{\bf v}\right ) - f\right ]
d^3{\bf x}d^3{\bf v}
\end{equation}

\noindent always decreases under arbitrary phase-space mixing.  Here,
$\overline{F}\left ({\bf x},\,{\bf v}\right )$ is the coarse-grained
distribution function.  Under a certain set of assumptions, the excess
mass function for a single cuspy halo with $\rho\propto r^{-\gamma}$
as $r\to 0$is $\propto f^{-2\left (3-\gamma\right )/\left
(6-\gamma\right )}$ as $f\to\infty$ where Thus, shallower cusps are
more mixed than steeper ones and merging cannot produce a cusp steeper
than the steepest cusp of the progenitors.  Moreover, $D(f)$ is
dominated by the contribution from the steepest cusp of the
progenitors in the limit $f\to \infty$ implying that the steepest cusp
survives. Dehnen's theorem is supported by numerical simulations that
follow the merger of equal-mass progenitors with various central
density laws \citep{kazantzidis}.

\subsection{Instability}
\label{sec:ROI}

Dehnen's theorem would seem to imply that the progenitors of
present-day NFW profiles have $r^{-1}$ density cusps since these are
preserved in the final product.  To understand why there should be
this primordial cusp profile} $r^{-1}$ we must explore what is missing
in Dehnen's arguments, namely collective effects such as those
associated with the ROI.  We find that the instability can lead to a
spontaneous change in the density profile and orbit structure of the
collapsing system by changing both the energies and angular momenta of
the particles in a coherent fashion. Although Dehnen's theorem does
not require these quantities to be conserved, it does not envisage
such globally coherent action.

The existence of the ROI was first established in the context of
anisotropic spherical equilibrium models \citep{fridman, merritt85}.
\citet{aguilar90} suggested that a similar instability was operating
in their collapsing systems.  \citet{pp87}, following \citet{ps81},
gave an analytic proof of the existence of the instability that
invokes a resonance in low angular momentum orbits such that the
orbital precession angle of a particle in one radial period is close
to $\pi$.  A review of the ROI can be found in \citet{me99}. 

It is interesting to note that the ROI is not the only instability
that may be operating during the process of halo formation.  The
self-similar solutions of \citet{fg84} and \citet{bert85} are unstable
to purely radial perturbations \citep{hw97}.  The instability is
driven by collective interactions between neighboring streams in the
phase-space flow and is similar to the instability in equilibrium
models found by \citet{henon}.  Once again an instability leads to
local variations in the energy of the particles which disrupt the
strict phase mixing sheets creating a coarse-grained distribution
function.

\section{NUMERICAL METHODS}
\subsection{Initial conditions}
\label{sec:IC}

For the spherically-averaged density profile of our initial
perturbation we use the mean density profile of a peak of the
primordial density field \citep{bbks, rg87}.  This density profile is
determined from the power spectrum $P(k)$ which in turn depends on
cosmological parameters.  We assume an Einstein-de Sitter cold dark
matter dominated cosmology with the power spectrurm

\begin{equation}
P(k) = P_{\rm CDM}(k)e^{-k^2l^2/2}
\label{eq:pofk}
\end{equation}

\noindent where $l$ is a smoothing scale \citep{rg87} and $P_{\rm
CDM}$ is from \citet{bbks}.  (While the current preferred cosmological
paradigm has changed significantly since the work of \citet{bbks} --
e.g. $\Lambda$CDM rather than Einstein-de Sitter -- our conclusions
hold irrespective of the cosmology.)  The initial density profile also
depends on the parameter $\nu$ where $\nu\sigma$ is the height of the
peak above the background.  In our simulations, we assume $l=100\,{\rm
kpc}$ and $\nu = 4$.

The simulation particles begin on a cubic grid and their positions
are perturbed radially so that the density matches the global density
profile $\langle \delta (r) \rangle$.  Velocities are started in the
Hubble flow.  The small-scale perturbations included in some of our
simulations are generated from Eq.\,\ref{eq:pofk} using the 
Zel'dovich approximation.

\subsection{N-body code}

Simulations are performed using an $N$-body code \citep{stiff03} that
is based on an oct-tree design and mulitpole expansion to compute
cell-cell interactions \citep{dehnen00}.  The code employs the $F_2$
softening kernel from \citet{dehnen01} and fixed time steps. The
opening angle is dynamically computed so that the maximum force error
is constant.

The softening length $\epsilon$ and time step $\Delta t$ are held
constant across our suite of simulations.  Our choice of softening
length is based on the criteria established in \citet{power03},

\begin{equation}
\label{eq:soft}
\epsilon ~\simeq~ \frac{4 R_{\rm vir}}{\sqrt{N_{\rm vir}}}~,
\end{equation}

\noindent where $N_{\rm vir}$ is the number of simulation particles
inside the virial radius $R_{\rm vir}$.  For our simulations (see
below) $R_{\rm vir} \sim 350$ kpc and $N_{\rm vir} \sim 5 \times 10^5$
yielding a softening length of 2 kpc.

We use a time step set by considering the maximum acceleration of the
particles at the first force calculation:
\begin{equation}
\Delta t = 0.1 \sqrt{ \frac{\epsilon}{a_{\mathrm{max}}}}~.
\end{equation}
\noindent
Approximately 9000 time steps are required to evolve the systems from
$z = 30$ to $z = 0$.


\subsection{Series of simulations}

We perform fives simulations of increasing  complexity.  Halos (a)
and (b) follow the collapse of a smooth, spherically symmetric density
perturbation.  Halo (a) is evolved with only radial forces while halo
(b) is evolved with the full three-dimensional force law.  That is, in
halo (a), the true force is replaced by its component in the radial
direction.

In halos (c), (d), and (e), we introduce small-scale perturbations.
Halo (c) is evolved with pure radial forces while halo (d) is evolved
with the true forces.  For halo (e) we include an external force meant
to mimic the effects of tidal fields from nearby galaxies.  The
potential which generates this field is taken to be
\begin{equation}
\Phi(r, \theta, \phi) \propto r^2 \sin^2 \theta \sin 2\phi.
\end{equation}

\section{Results}

\subsection{Global properties of the simulated halos}

Table \ref{tab:results} provides global properties of the 
simulated halos.  The table includes
the mass $M_{200}$ contained within the virial radius $R_{200}$,
defined to be the radius which encloses an average density that is 200
times the background value, $\rho_b = 3 H^2 / 8 \pi G$.  Also given is
the ``alternative'' spin parameter defined by \citet{bull01}:
\begin{equation}
\label{eq:lambda}
\lambda^{'} ~\equiv~ \frac{J}{\sqrt{2} M V R}~,
\end{equation}
where $J$ is the total angular momentum, $M$ is the total mass, and
$V$ is the circular velocity at radius $R$.  This 
definition of the spin
parameter is equivalent to the usual form, $\lambda = J |E|^{1/2} / G
M^{5/2}$, when measured for a singular isothermal sphere at the virial
radius and is much easier to compute for simulated halos.  In
equation (\ref{eq:lambda}) $R,\,M,\,$ and $V$ are taken to be
$R_{200},\,M_{200}$, and $V_{200}$ respectively.  Table
\ref{tab:results} also gives the mean velocity anisotropy parameter
for each halo.

\begin{equation}
\label{eq:beta}
\beta = 1 - \frac{\sigma^2_t}{2\sigma^2_r}.
\end{equation}

\noindent where $\sigma_t$ ($\sigma_r$) is the rms tangential (radial)
velocity within the virial radius.

While the virial masses and radii are very similar across the sequence
of simulations, $\lambda^{'}$ and $\beta$ vary greatly.  In the cases
where the particle accelerations are constrained to be radial
(simulations (a) and (c)), $\beta \simeq 1$.  Non-radial forces lead
to a small $\lambda'$ even with smooth spherically symmetric initial
conditions.  Essentially, non-radial forces drive the system to a
prolate-triaxial structure and torques between the inner virialized
region and the outer region generate a small amount of bulk angular
momentum.  The effect is enhanced by more than an order of magnitude
when perturbations are included in the initial conditions.  Of course,
the addition of an external quadrupole field, with a suitably chosen
amplitude, leads to bulk angular momentum comparable to what is found
in full cosmological simulations.

\begin{figure}
\epsscale{1.0}
\plotone{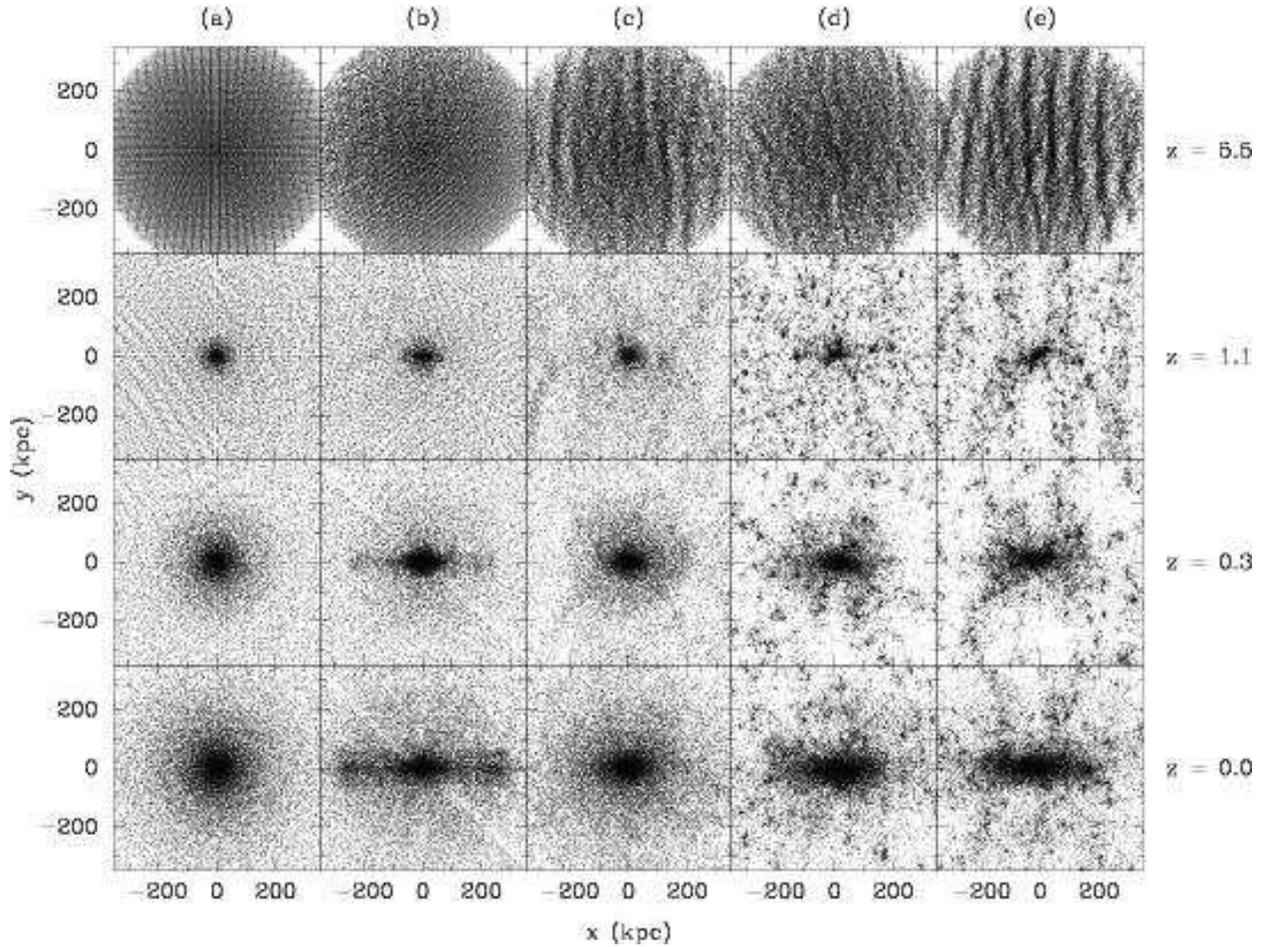}
\caption{Snapshots of the simulations at various redshifts: from top
to bottom, $z = 5.5$, 1.1, 0.3, and 0.  The simulations are shown
projected in the $x-y$ plane and rotated so that their longest axis is
along the $x$-axis. The columns correspond to the halo labels
presented in Table \ref{table:params}.}
\label{fig:sims}
\end{figure}

Snapshots of the halos at various stages in their evolution are shown
in Figure \ref{fig:sims}.  As expected, halo (a) maintains nearly
perfect spherical symmetry whereas (b) develops into a strongly
prolate system.  This change in shape is the result of the ROI.
Likewise, halo (c), though beginning from slightly aspherical initial
conditions, develops into a relatively smooth and only mildly prolate
system.  Halo (d), on the other hand, has significant substructure and
is more strongly prolate.  Halo (e) is very similar in appearance to
halo (d).

\subsection{Density profiles}

\begin{figure}
\epsscale{1.0}
\plotone{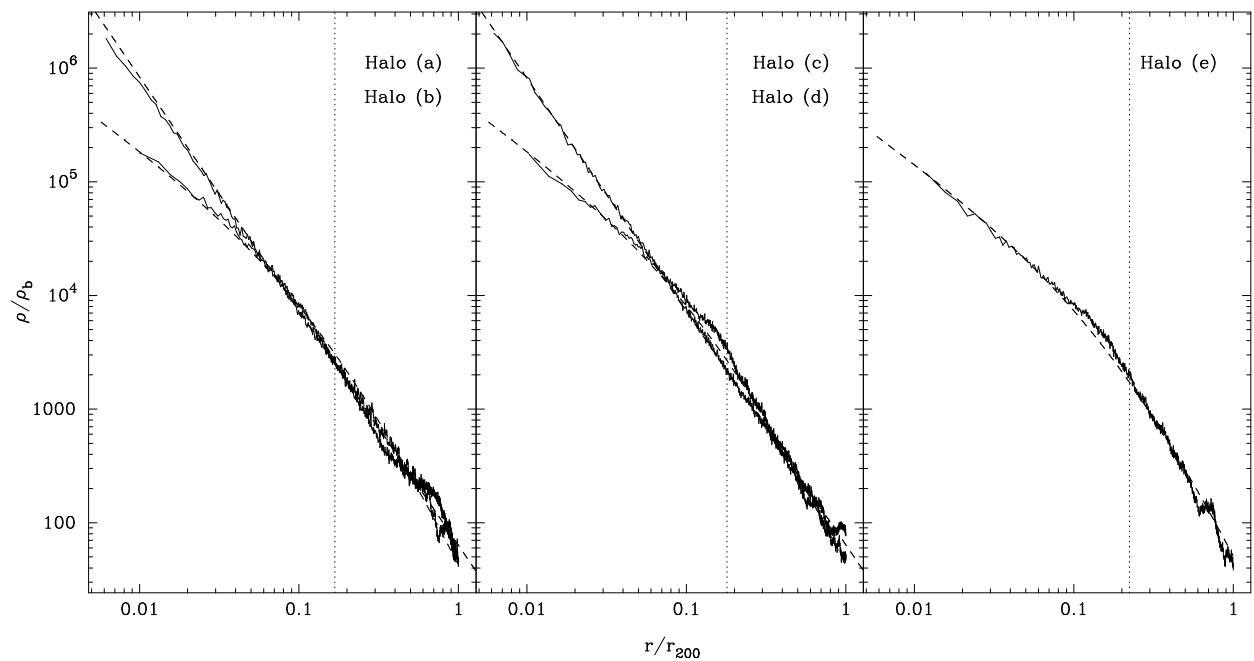}
\caption{Density profiles of each of the final halos at $z = 0$ (solid
lines).  Also shown (dashed lines) is
the best-fit NFW density profile for halos (b), (d), and (e) and
power-law fits for halos (a) and (c).}
\label{fig:density}
\end{figure}

The density profiles for our halos are shown in Figure
\ref{fig:density}.  To construct the profiles, we use radial bins with
equal particle number (typically 500).  The difference between the
density profiles for halos with pure radial forces and those which do
not is obvious -- the steep inner cusp of the former becomes much
shallower when the ROI is allowed to proceed.  Our results agree with
those of \cite{hjs} in that an NFW-like density profile can be
generated from a smooth collapse without appealing to merging.

The steep density profiles of halos (a) and (c) are well fit by a
power-law, $\rho \propto r^{\alpha}$.  Least-square fits to the data
give $\alpha = -2.022 \pm 0.004$ for halo (a) and $\alpha = -2.050 \pm
0.003$ for halo (c).  These profiles are consistent with results from
the semi-analytic self-similar models which predict, for pure radial
collapse, a final density profile of $\rho \propto r^{-\mu}$ with
$\mu\simeq 2$.

The other halo profiles shown in Figure (\ref{fig:density}) are fit
with an NFW profile (equation \ref{eq:nfw}).  For Halo (b), the best
fit profile has a concentration parameter $c_{200} = R_{200} / r_s =
6.0$, while for Halo (d) we find $c_{200} = 5.6$ provides the best
fit.  The concentration parameter for Halo (e) is $c_{200} = 4.5$.
These concentrations are lower, by a factor $\sim 2$, than those found
by \citet{nfw} for halos of a similar mass.  Also, the density
profiles in our halos show a sharper transition from $\rho \sim
r^{-3}$ in the outer regions to $\rho \sim r^{-1}$ in the inner core.
This seems to be a common feature of halos that have undergone the
ROI and is present in other simulations not shown here.

\begin{figure}
\epsscale{1.0}
\plotone{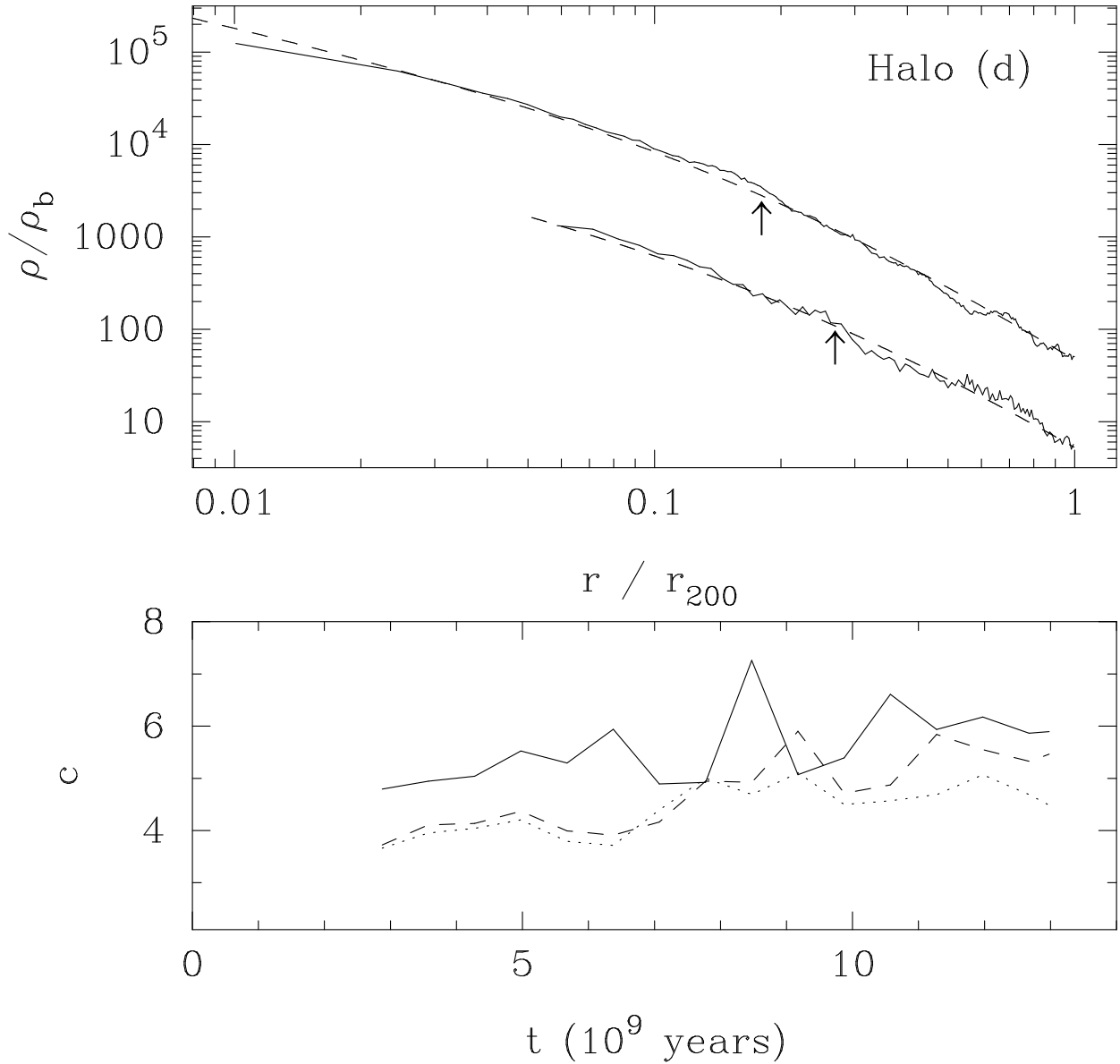}
\caption{(Top panel) The density profile of halo (d) at early ($z
=1.72$, dashed line) and late ($z = 0$, solid line) times.  The former
is offset by 1 dex.  Arrows indicate the NFW scale-length $r_s$.
(Bottom panel) The concentration parameter of the
NFW fit to density profiles for the halos as they evolve in time.  The
solid line indicates halo (b), the dashed line is halo (d), and the
dotted line is halo (e). }
\label{fig:nfw}
\end{figure}

The density profiles of the halos that undergo the ROI rapidly and
spontaneously develop the NFW shape.  In fact, once the number of
particles within the virial radius is large enough to calculate an
accurate density profile (Figure \ref{fig:nfw}, top panel), the
profile is already well fit by NFW formula.  However, the
concentration parameter of the halo stays relatively constant (Figure
\ref{fig:nfw}, bottom panel) throughout the lifetime of the system
even though the virial radius increases steadily.  The implication is
that the system is evolving in a self-similar fashion.  The ROI drives
a transition in the inner regions from the self-similar radial infall
solutions to a stable state with isotropic velocities and a softer
cusp.  The system in this region is reminiscent of the steady-state
self-similar solutions discussed in \citet{evans, hw95}.

In this connection we note that Henriksen and Widrow (1995)
found a steady-state scale-free solution in spherical symmetry with
velocity isotropy that has a density profile $\propto r^{-2/\delta}$
with $\delta>1$.


\subsection{Angular momentum distribution}

As with the density profile, the specific angular momentum
distribution also has a universal profile.  \citet{bull01} performed a
high-resolution cosmological simulations and calculated the mass
$M(<j_z)$ with angular momentum less than $j_z$ for individual halos.
They found $M(<j_z)$ was well fit by the simple formula,

\begin{equation}
\label{eq:bullock}
M(<j_z) = \left\{\begin{array}{ll}
M_{\rm{tot}} \frac{\mu j_z}{j_0 + j_z} & \mbox{if $j<j_{\rm max}$}\\
M_{\rm{tot}} & \mbox{otherwise}~,\end{array}\right.
\end{equation}

\noindent where $\mu$ and $j_0$ are constants and $j_{\rm max}\equiv
j_0/\left (\mu-1\right )$.  \citet{bull01} also found that the angular
momentum in their halos is well aligned; only a small fraction (about
5\%) of the halo mass was counter-rotating (i.e., $j_z<0$).  A similar
analysis by \citet{chen02} found that in about half of their halos,
greater than 10\% of the halo mass had $j_z<0$.

\begin{figure}
\epsscale{1.0}
\plotone{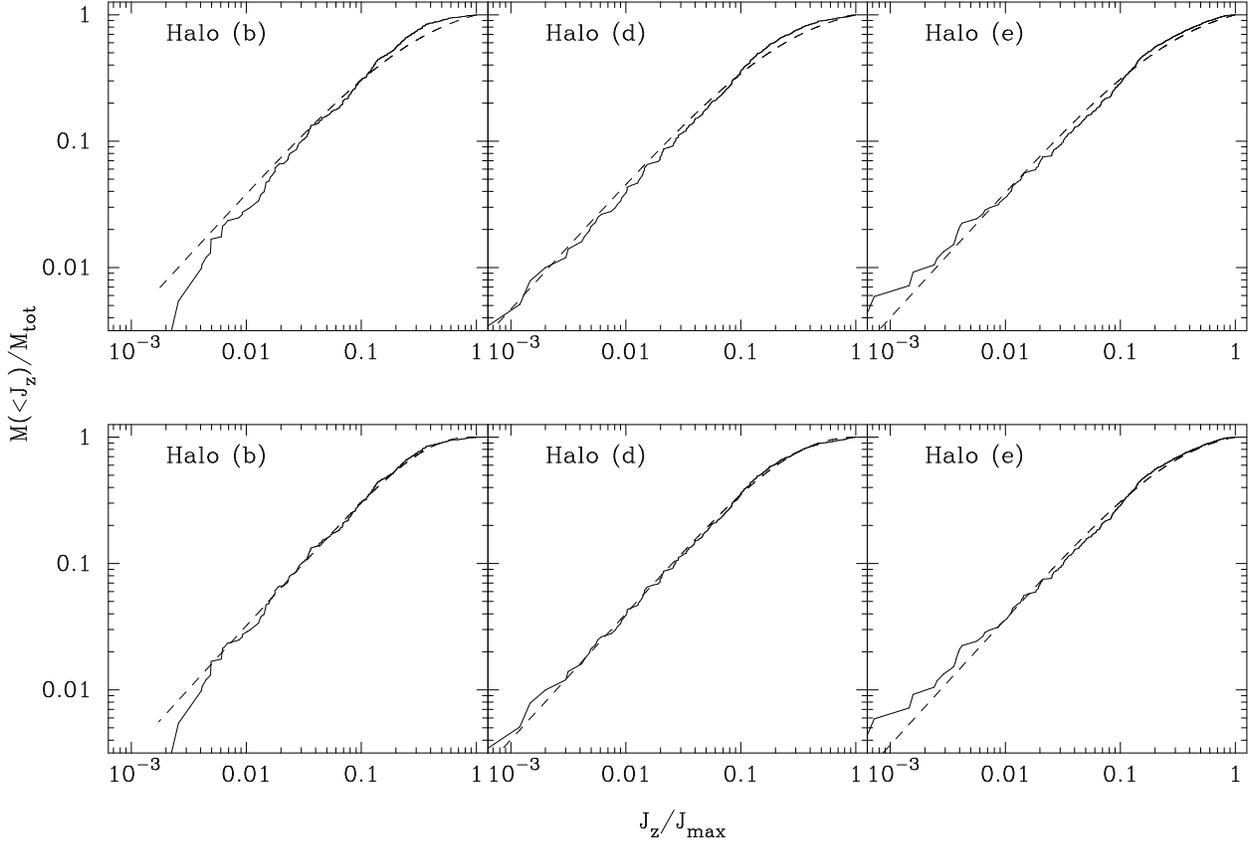}
\caption{The angular momentum distribution for three final halos at $z
= 0$.  The distribution was calculated as in \citet{bull01}, and shows
the total mass $M(<j_z)$ that has angular momentum less than $j_z$.
(Top panel) Fits to the formula of \citet{bull01} (dashed lines).  The
shape parameter for each halos is $\mu = 1.34$ for halo (b), $\mu =
1.27$ for halo (d), and $\mu = 1.33$ for halo (e). (Bottom panel) Fits
to our formula (equation \ref{eq:jfit}.  For halo (b), $\mu = 1.45$
and $\Gamma = 1.52$; for halo (d), $\mu = 1.33$ and $\Gamma = 1.32$; and
for halo (e), $\mu = 1.37$ and $\Gamma = 1.14$.}
\label{fig:bullock}
\end{figure}

Following \citet{bull01}, we calculate the angular momentum
distribution in our halos.  In Figure \ref{fig:bullock} (top panel),
we show $M(<j_z)$ along with the least-squares best-fit to equation
(\ref{eq:bullock}).  In general, the trends seen in the halos of
\citet{bull01} are also present here; $M(<j_z) \propto j_z$ for $j_z
\ll j_{\mathrm{max}}$, with a smooth turnover near $j_z \sim j_
{\mathrm{max}}$.  However, equation (\ref{eq:bullock}) does not fit
this turn over well, especially for the low-$\lambda$ halos (b) and
(d), where it underestimates the mass $M(<j_z)$.  The angular momentum
is also not well aligned within these two halos; about half of their
mass is anti-aligned and was excluded during the construction of their
angular momentum profiles.  Halo (e), which was evolved with an
external quadrupolar force, has a spin parameter much larger than the
other two halos. Only a small fraction of the mass (less than 5\%) in
this halo is counter-rotating and the fit to equation \ref{eq:bullock}
is significantly better than in the other cases.

To further illustrate this point, we consider the following 
generalization of equation \ref{eq:bullock}

\begin{equation}
\label{eq:jfit}
M(<j_z) = M_{\mathrm{tot}} \frac{\mu j_z / j_0}
{\left (1 + (j_z / j_0)^{\Gamma}\right )^{1/\Gamma}},
\end{equation}
where $\Gamma$ controls the ``sharpness'' of the transition from the
power-law.  Figure \ref{fig:bullock} (bottom panel) shows our halo
angular momentum profiles and the best fits to this new formula with
$\Gamma = 1.52/1.32/1.14$ for halos (b), (d), and (e) respectively.
The implication is that the fitting formula from \citet{bull01} is
most appropriate for cosmological halos that have a reasonable spin
parameter and angular momentum generated by an external tidal field.


\subsection{Velocity Dispersions}

\begin{figure}
\epsscale{1.0}
\plotone{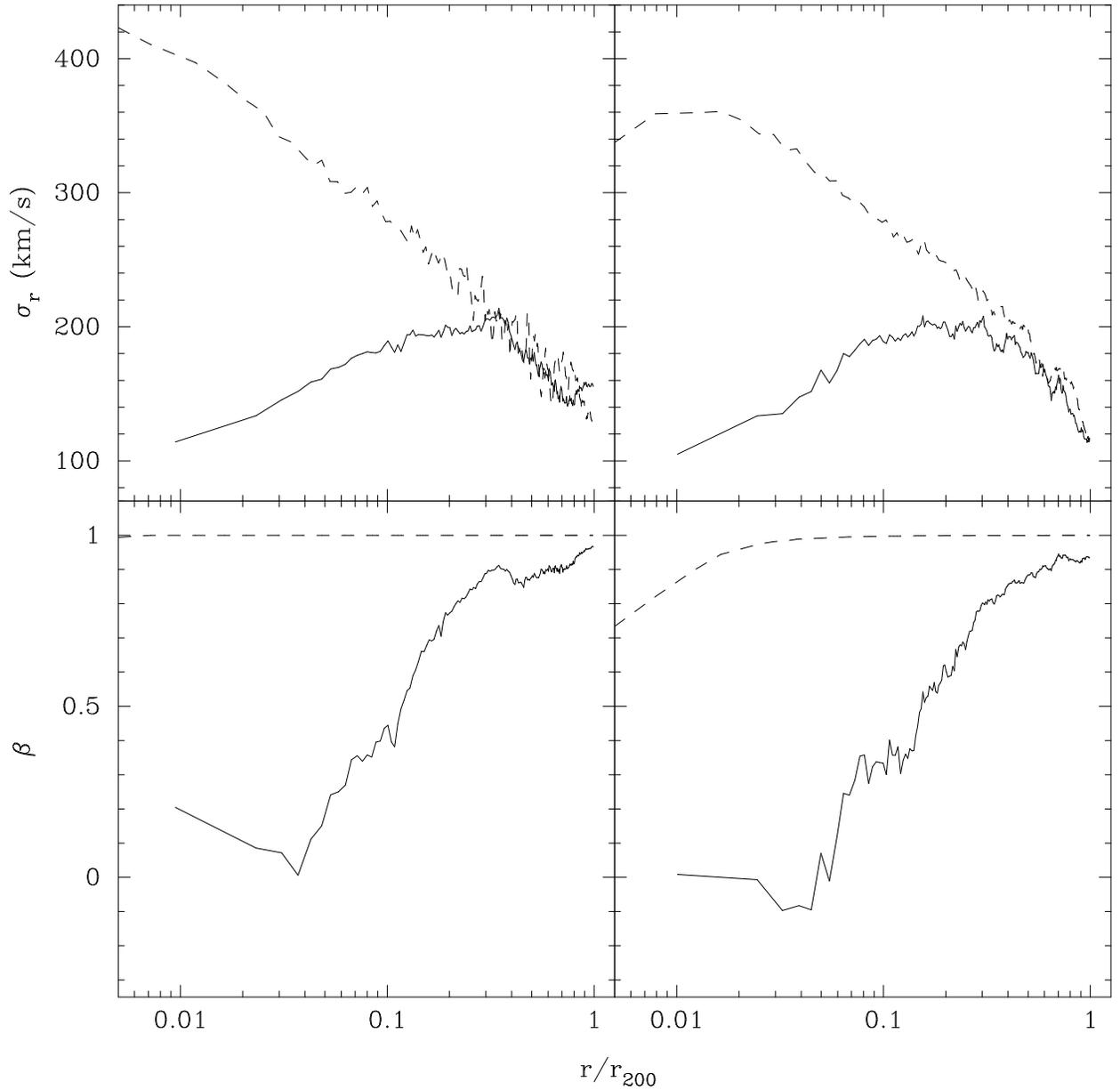}
\caption{Velocity dispersions, in radial bins, for the final halos.
Shown is the radial velocity dispersion $\sigma_r$ (top panels) and
the velocity anisotropy parameter $\beta = 1 - \sigma_t^2 /
\sigma_r^2$ (bottom panels).  The halos with constrained radial
accelerations are the upper velocity dispersions in the top panels,
and have $\beta \sim 1$ in the bottom panels.}
\label{fig:sigma}
\end{figure}

In general, the orbits of dark matter particles in simulated halos are
isotropic near their centres and more radial further out
\citep{colin00, fm01}.  To explore the connection between this trend
and the ROI, we calculate velocity dispersion profiles for our halos.
The results are shown in Figure \ref{fig:sigma}.  The difference
between the profiles for halos evolved with radially constrained
orbits and those that have undergone the ROI is striking.  The latter
exhibit a turnover in $\sigma_r$ near $r_s$ whereas the former have a
velocity dispersion that increases monotonically as $r\to 0$.  We also
show the anisotropy parameter $\beta$ (equation \ref{eq:beta}) as a
function of radius.  For models (b) and (d), $\beta$ increases from
$\sim 0$ at small $r$ to $\sim 0.9$ at large $r$ with the transition
occurring at $r\simeq r_s$.  The trend is similar to what is found in
cosmological simulations although there, $\beta$ never gets much above
$0.5$.

It is worth noting that the $\sigma_r$ profiles for halos $(a)$ and
$(c)$ are not pure power-laws in $r$ contrary to what one might expect
given that these models evolve from scale-invariant initial conditions
and yield power-law density profiles.  Consider, for example,
a spherical halo with
pure radial orbits and density law

\be
\label{eq:rminus2}
\rho(r) = \rho_0
\left\{\begin{array}{ll}
\rho_0 r_0^2/r^2 & \mbox{if $r<r_0$}\\
0 & \mbox{otherwise}~,\end{array}\right.
\ee

\noindent This system provides an idealized model for halos (a) and
(c).  The distribution function for this system is given by $f\left
(E,\,j\right ) = F_0\Theta\left (E \right ) E^{-1/2} \delta \left
(j^2\right )$ where $\Theta$ is the Heaviside function $F_0$ is a
constant \citep{fridman}.  This form for the distribution function
assumes that the potential for $r<r_0$ is $\psi = 4\pi G\rho_0
r_0^2\ln{r/r_0}$.  For this distribution function, $\sigma_r^2 =
\psi(r)/2$ and is therefore not a power law.

\subsection{Phase-Space Density}

\citet{taylor01}, and more recently \citet{dehnen05}, investigated the
``phase-space density'' of dark matter halos in cosmological
simulations.  More precisely, the computed the ratio $\rho/\sigma^3$ as
a proxy for the phase-space density.  They found that this quantity is
universally a power-law,
\begin{equation}
\frac{\rho}{\sigma^3} \propto r^{- \alpha},
\end{equation}
with $\alpha \simeq 1.84$ when considering the total velocity
dispersion and $\alpha \simeq 1.92$ when just the radial velocity
dispersion is used.

\begin{figure}
\epsscale{1.0}
\plotone{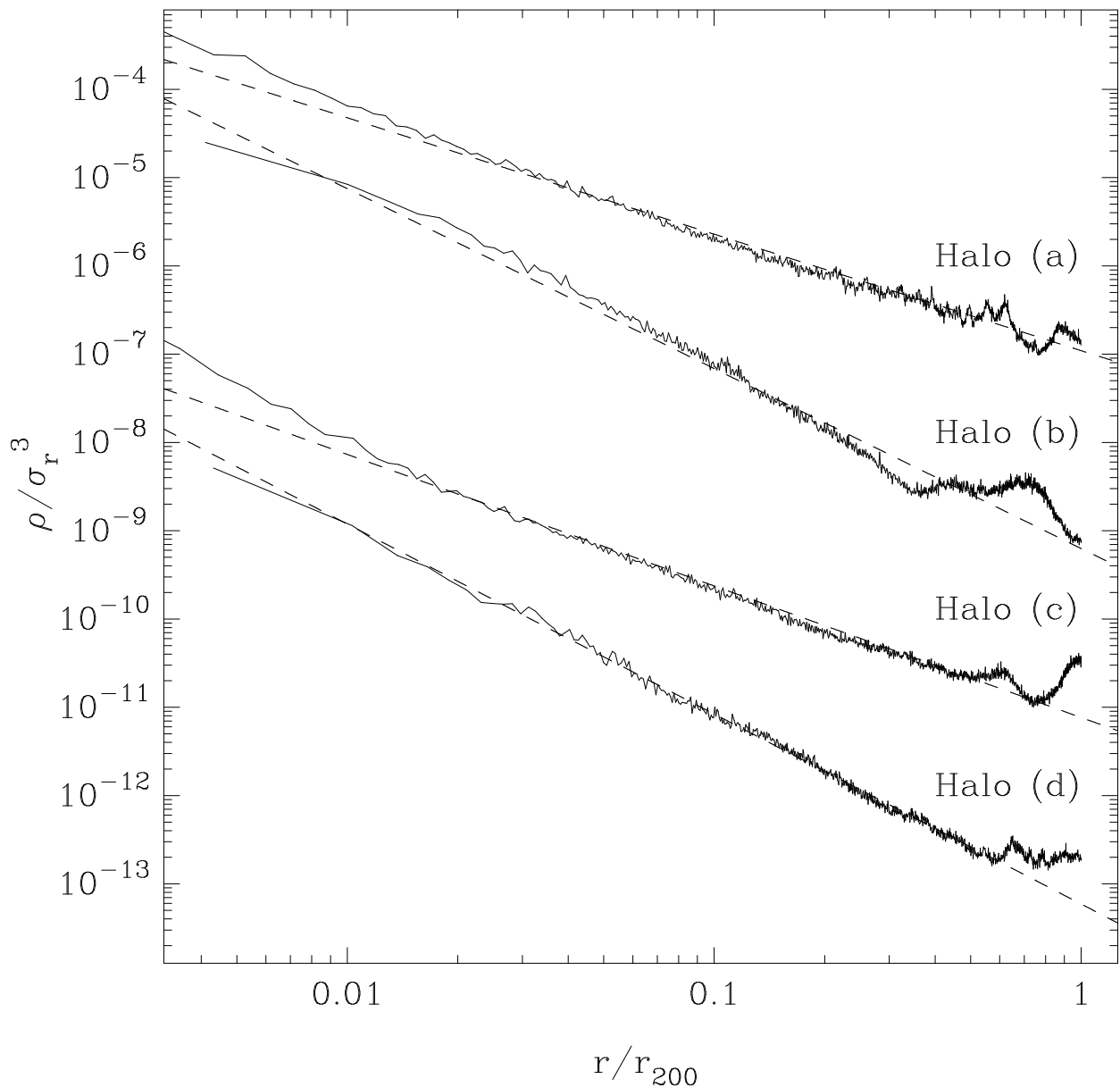}
\caption{The phase-space density of the final halos, $\rho /
\sigma_r^3$ (solid lines), along with their best fit slopes (dashed
lines).  Curves are off-set by 1 dex for clarity.
Fits are as follows: Halo (a) -- $\alpha = -1.32$.  Halo (b)
-- $\alpha = -2.04$.  Halo (c) -- $\alpha = -1.49$.  Halo (d) --
$\alpha = -2.15$.  Halo (e) -- $\alpha = -2.04$.}
\label{fig:phase}
\end{figure}

In Figure \ref{fig:phase}, we show $\rho / \sigma_r^3$ for halos
(a)-(d).  The curve for halo (d) is closest to a pure power-law.  The
result seems rather curious since it is halos (a) and (c) that exhibit
pure power-law density profiles.  However as pointed out in the
previous section, $\sigma_r$ is not a power law and neither will be
$\rho/\sigma_r^3$.  Evidently, the ROI changes both $\rho$ and
$\sigma_r$ in just such a way so as to yield a power law for the ratio
$\rho/\sigma_r^3$.  Whether this result is accidental or telling
us something deep about the ROI remains an open question.


\section{THE RADIAL ORBIT INSTABILITY}

The most striking feature of the final halos (b), (d), and (e) is the
strongly prolate shape, a generic feature of the ROI.  To further
explore the development of the ROI we calculate the bar-strength
parameter, defined as the relative amplitude of the bisymmetric ($m =
2$) Fourier component of the mass density averaged over some region of
interest.  Operationally, this quantity is calculated through the
expression \citep{shen04}:
\begin{equation}
A(t) = \frac{1}{N} \left| \sum_{j} \exp(2 i \phi_j ) \right|,
\end{equation}
where $\phi_j$ is the azimuthal coordinate of the $j$th particle in
the sum, and $N$ is the total number of particles considered.  The
evolution of the bar strength for halos (b), (d), and (e) is shown in
Figure \ref{fig:bar}.  In the top panel, $A(t)$ is calculated for all
particles within the virial radius $R_{200}(t)$.  For this choice of
region, $A(t)$ is relatively constant implying that the region
undergoing the instability scales roughly with the virial radius and
that the strength of the bar inside the virial radius is roughly
constant in time.  In the bottom panel $A(t)$ is calculated using all
particles in the simulation and we find that $A(t) \propto t^{\alpha}$
where $\alpha = 2.6, 2.0,$ and $1.6$ for halos (b), (d), and (e)
respectively.  Perhaps more telling is the observation that the final
bar strength within the virial radius is the same in each of these
runs.  The implication is that so long as the initial conditions allow
for roughly spherical accretion, the ROI drives systems to a common
(universal) final state.  The ROI develops more rapidly in systems
that are initially more strongly biased to radial orbits such as halo
(b).  But the final state is, at least in the context of the
bar-strength, independent of initial conditions.

\begin{figure}
\epsscale{1.0}
\plotone{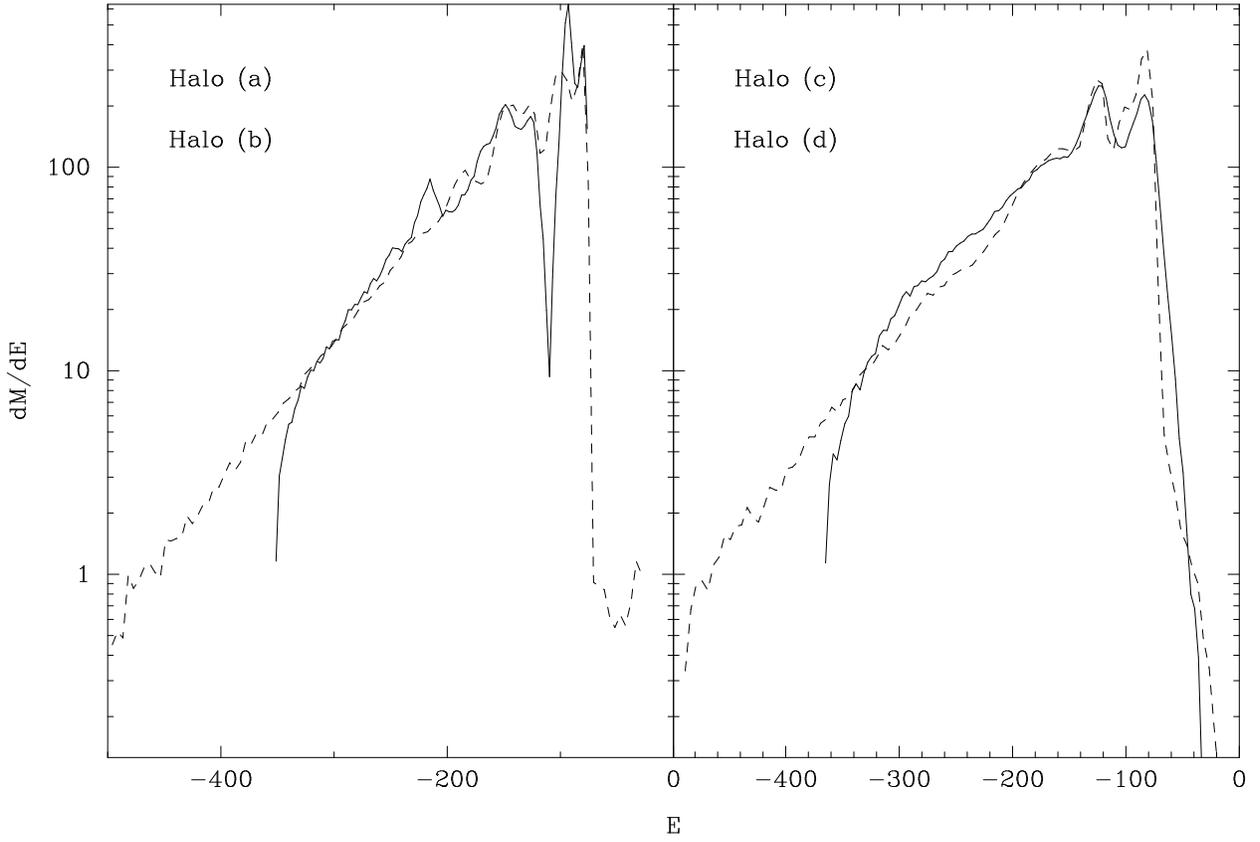}
\caption{The differential energy distribution, $\dd M / \dd E$ for
each of the halos.  Dashed lines indicate those halos (a and b) which
do not undergo the radial orbit instability.}
\label{fig:dmde}
\end{figure}

Further evidence of the dramatic effect of the ROI can be seen in the
energy and angular momentum distributions of particles.  Figure
\ref{fig:dmde} shows the differential energy distribution $\dd M/\dd
E$ for halos (a)-(d).  (The result for (e) is similar to that for (b)
and (d).)  The ROI appears to introduce a cut-off in the energy
distribution at large binding energies.  \citet{bt87} point out the
$\dd M /\dd E$ is primarily determined by the spherically-averaged
density profile.  Our results complement this rule in that they
provide an example where a change in $\dd M/\dd E$ goes hand-in-hand
with a change in $\rho$.

As previously discussed, the ROI leads to a change in the inner
structure of dark matter halos from $r^{-2}$ with radial orbits to
$r^{-1}$ with isotropic orbits.  We note that while the former
resembles the self-similar state discussed by \citet{fg84} and
\citet{bert85} the latter resembles the self-similar state found in
\citet{hw95}.  Moreover, dimensional analysis suggests that the rms
angular momentum of particles in the inner region scales with radius
as $j\propto \sqrt{GrM(r)}$.  This scaling is precisely the relation
required to preserve self-similarity, as noted by \citet{sikivie97},
\citet{nusser01}, \citet{ledelliou03}, and \citet{asc04}.  Here, we
argue that this rms angular momentum is generated by torques due to
the evolving bar-like perturbation.  

\begin{figure}
\epsscale{1.0}
\plotone{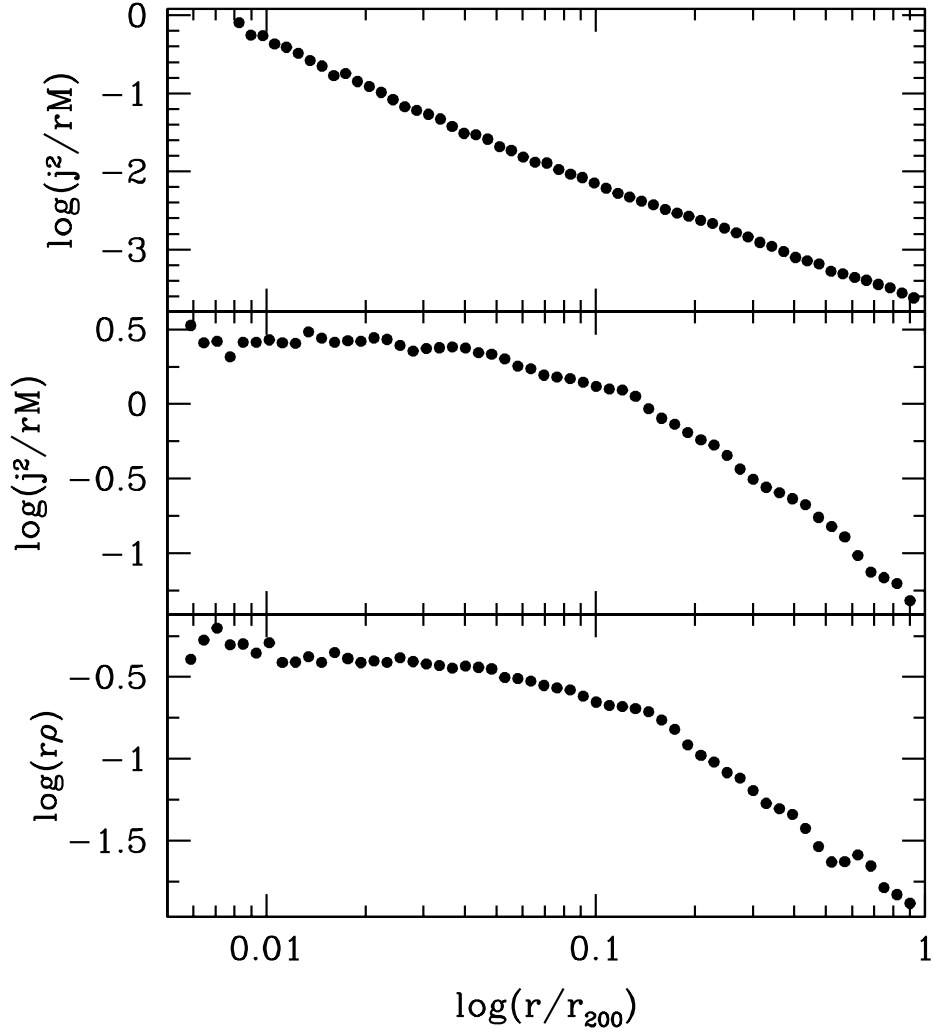}
\caption{Comparison of angular momentum and density profiles.
$j^2/rM(r)$ as a function of $r$
for halo (c) (top panel) and halo (d) (middle panel).
$\rho(r) r$ for halo (d) (bottom panel).}
\label{fig:j2}
\end{figure}

A striking confirmation of the connection between the density and
angular momentum profiles is given in Figure \ref{fig:j2}.  The figure
shows $j^2/rM(r)$ as a function of $r$ for halos (c) and (d) as well
as $r\rho(r)$ for halo (d).  Since halo (c) is run without
non-radial forces, the angular momentum distribution is given by
initial conditions.  Let $r_i$ be the initial radius of a particle and
$v_{\perp,i}$ be its initial transverse velocity.  On average,
$v_{\perp,i}$ will be independent of $r_i$.  Moreover, the relation
between $r_i$ and the final average radius of a particle is given,
by the relation 

\be
\frac{4\pi}{3}r_i^3\rho_i =  M(r).
\ee

\noindent For an $r^{-2}$ density law, $M(r)\propto r$ and therefore
$j\propto r_iv_{\perp,i}\propto r^{1/3}$ in reasonable agreement with
the result for halo (c).

Turning to halo (d), we see that $j^2$ evolves to the value predicted by
dimensional analysis precisely in the  region of the NFW inner
cusp.  That is, $r\rho(r)$ and $j^2/rM(r)$ are constant over the same
range in $r$.  Indeed, for halo (d), these two curves look remarkably
similar.

We conclude this section with a brief discussion of the underlying
physics of the ROI.  The literature on this subject was reviewed in
\citet{me99}.  In spherical systems, highly elongated orbits are
nearly closed and therefore act like a precessing ellipse or rod.  In
a system dominated by such orbits, a bar-like disturbance grows since
orbits precessing slowly across the perturbation are trapped.

For our simulations this argument must work for radial orbits having
an initial $r^{-2}$ density profile with the orbit distribution of
halos (a) or (c), i.e., models that are unstable to the ROI. 
Recall that the azimuthal angle for a particle travelling from
pericenter $r_1$ to apocenter $r_2$ and back changes by

\be
\Delta \phi = 2j\int_{r_1}^{r_2}
\frac{dr'/r'^2}{\left (2\left (\psi(r')-{\cal E}\right ) 
- j^2/r'^2\right )^{1/2}}~.
\ee

\begin{figure}
\epsscale{1.0}
\plotone{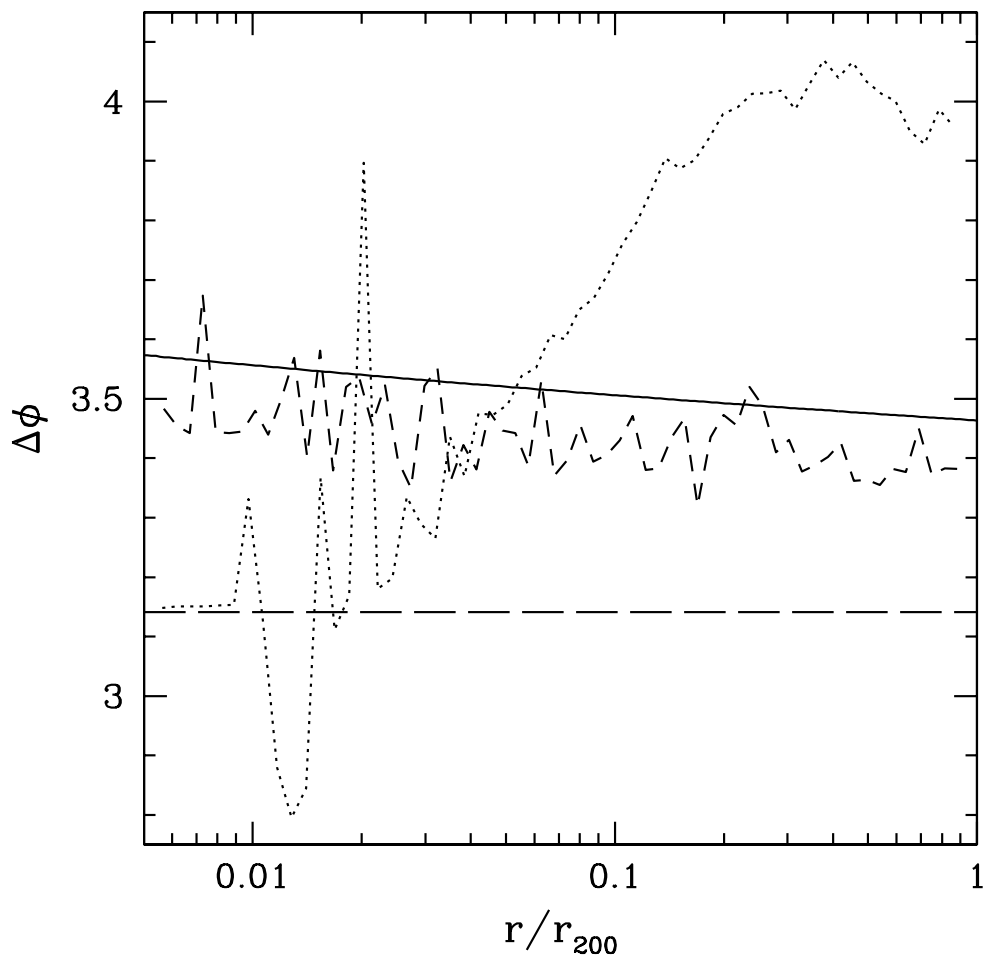}
\caption{Change in azimuthal angle for one radial period as a function
of $r$.  In general, $\Delta \phi$ depends on the energy and angular
momentum of the particle.  Here, we replace the energy with angular
momentum with their values averaged over radial bins of radius $r$.
Halo (c) -- dashed line; Halo (d) -- dotted line; Semi-analytic
calculation for halo (c) -- solid line.  A value of $\Delta\phi = \pi$
(long-dashed line) corresponds to a closed orbit where the particle
makes two radial oscillations per orbit.}

\label{fig:deltaphi}
\end{figure}

\noindent $\Delta \phi$ is a function of ${\cal E}$ and $j$ but can be
reduced to a function of $r$ by replacing ${\cal E}$ and $j$ with
their rms values in spherical shells.  The results for halos (c) and
(d) are calculated numerically and shown in Figure \ref{fig:deltaphi}.

Before discussing the significance of this result we check the
numerical result by calculating $\Delta \phi$ semi-analytically in an
$r^{-2}$ halo.  In particular, we assume the density to be given by
equation \ref{eq:rminus2}.  We take the average energy to be
$\psi(r)/2$ which is the exact result assuming purely radial orbits.
(One proves this result by using the distribution function discussed
above, but it is to be expected in a virialized system).  As for the
angular momentum, we take $\langle j(r)\rangle = j(r_0)\left
(r/r_0\right )^{1/3}$ where $j(r_0)$ is calculated from the
simulations.  With the change of variables $r'\to \sqrt{r_0 r}/u$ we
have

\be
\Delta \phi = 2j\int_{u_2}^{u_1}\frac{du}{\left (
\eta\log{u} - u^2\right )}~.
\ee

\noindent where $\eta\equiv 2M(r_0)r_0/j_0^2$.  The results of this
integral are also shown in Figure \ref{fig:deltaphi} and agree
reasonably well with those for halo (c).

We propose that the key result is the near constancy of $\Delta \phi$
at a value close to $\pi$ for halo (c).  By contrast, $\Delta \phi$
varies considerably with radius for halo (d).  Recall that halo
(c) {\it is} unstable to the ROI whereas halo (d) has already
undergone the ROI.  Suppose that there is a perturbation along a particular
axis, say, a subhalo falling in on a particular radial orbit.  Particles
at an angle $2\left (\Delta \phi - \pi\right )$ from this axis but
over a large range in radius will precess into the perturbation and
act in concert to enhance it.  

Together the preceding considerations make explicit the criteria
given in \citet{pp87} for the ROI. The bias towards radial orbits is
evident for halo (c), but in addition we have shown that the resonance
condition that these authors give (esentially that $\Delta\theta\equiv
2\left (\Delta \phi - \pi\right )$ is small) can hold simultaneously
at many radii.  The mechanism may, in fact, be analogous to the bar
instability in rotating disks (See \citet{bt87} and references
therein).

The simultaneous development of the $j^2/rM(r)$ $= constant $ state
shown in figure (\ref{fig:j2}) for halo (d) may also be roughly
understood in terms of these ideas. Given a bar-like perturbation that
has $M_p(r)=f(r)M(r)$ and noting that particles within $\Delta\theta$
of this radius will be most affected by the perturbation during the
next two radial orbits , we can expect them to acquire an rms $j$
proportional to the torque multiplied by the dynamical time namely \be
j~\propto~ \frac{GfM}{r^2}\times r\times \sqrt{\frac{r^3}{GM}}.  \ee
This yields $j\propto f(r)\sqrt{Mr}$. Thus if $f$ is constant, as may
be expected from our initial conditions, we obtain what is observed in
the simulation. However this begs the question of universality if
$f(r)$ were some perverse function. We suspect that universality may
nevertheless hold because of the rapid growth of the bar perturbation
and the nearly simultaneous development of $j^2$. Resonating particles
will be most strongly torqued towards the end of their precession, and
hence at late dynamical time. By then the bar contains a large
fraction of the system mass inside the virial radius (Figure
\ref{fig:bar}, top panel) and the initial $f(r)$ is likely to be
forgotten.

\begin{figure}
\epsscale{1.0}
\plotone{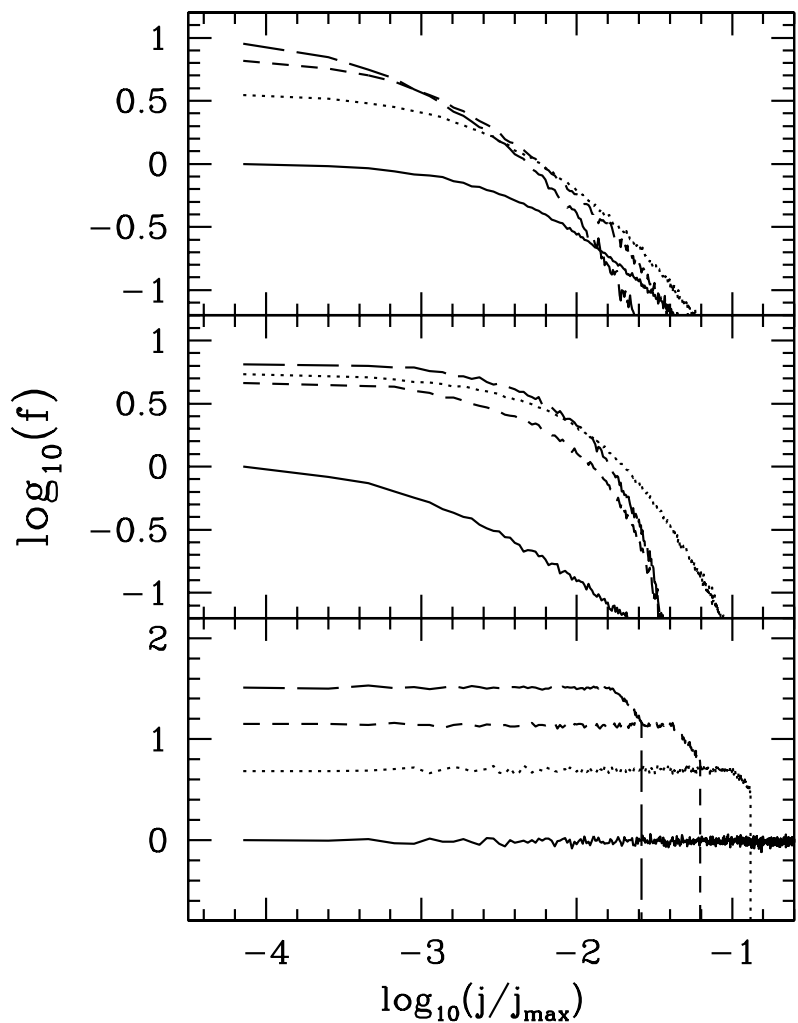}
\caption{Distribution function as a function of $j$ for different
values of the energy.  Top panel -- halo (c); Middle panel -- halo (d);
Bottom panel -- distribution function recovered from an N-body
realization of an analytic model with $f = f(E)$, i.e., no $j$ 
dependence.  Line types correspond to different energies.  In
order of increasing binding energy, they are solid, dashed, dotted
and long-dashed.}
\label{fig:fofej}
\end{figure}

We conclude our discussion of the ROI by considering the phase-space
distribution function, $f\left ({\bf x},\,{\bf v}\right )$, for
various halos.  The criteria for the development of the ROI from
\citet{pp87} is that the $f$ diverges as $j\to 0$.  For a spherically
symmetric system in equilibrium, $f$ is a function of the energy and
angular momentum.  Though our halos are aspherical and evolving, an
approximate 2-integral distribution function can be calculated by
spherical averaging and assuming that the system is in equilibrium.
$f$ is then given by the mass distribution in terms of the energy and
angular momentum ($d^2M/dE\, dj$) divided by the radial period, $T_r$
(see, for example, \citet{bt87}).  In Figure \ref{fig:fofej}, we show
the results for $f$ as a function of $j$ at various values of $E$.
Once again, the distinction is clear: halo (c) shows evidence of a
divergence in $f$ at small $j$ suggesting a susceptibility to the ROI
while halo (d) has a distribution function that is flatter at small
$j$ indicating that the ROI has already operated and driven the system
to a more stable state.


\section{CONCLUSIONS}

In the hierarchical clustering scenario, halos grow through the
accretion of smaller systems and, occasionally, major merger events
involving comparably-sized objects.  Despite the apparent random
nature of this process, halos exhibit a number of ``universal''
characteristics across many orders of magnitude in mass.

The accretion phase of halo formation is reminiscent of spherical,
self-similar collapse.  As illustrated in \citet{hjs} and in the
present work, the missing ingredient appears to be the ROI.  In
effect, this collective phenomena prevents halos from forming with
cusps much steeper than $r^{-1}$.  Our analysis of the energy and
angular momentum distribution provides strong additional evidence of the
importance of the ROI in shaping the structure of dark halos.

Recently, \citet{lu} considered a simple, phenomenological model for
halo growth that reproduced many features of simulated dark halos.  In
the model, halo growth takes place in two phases, a fast accretion
phase in which orbits are isotropized, and a slow accretion phase in
which orbits remain more or less radial.  The first phase leads to the
$r^{-1}$ cusp while the second produces the $r^{-3}$ outer halo.  In
effect, the ROI instability may be the physical mechanism for
isotropization of orbits in the first phase of their scenario. We have
argued that the merger theorem of \citet{dehnen05a} together with
simulations by \citet{kazantzidis} suggest the need for softening of
the central density cusp by some coherent process. Thus the ROI is a
candidate for the basic collective relaxation mechanism that takes
steep self-similar infall cusps to the inner NFW behaviour.

The bulk angular momentum distribution of \citet{bull01} appears to be
a natural corollary of this picture in that a quadrupole field applied
to the bar-like structure that develops from the ROI yields an $M\left
(<j_z\right )$ given approximately by equation
\ref{eq:bullock}. Moreover the rms angular momentum produced by the
ROI alone is given by the dimensional argument discussed in the text
and is not in fact too dissimilar from the bulk distribution produced
tidally, although the amplitude is much smaller

Perhaps the most curious result is that the ROI drives the combination
$\rho/\sigma^3$ toward a power law in $r$ even as it drives $\rho$ away
from a pure power law.  Whether $\rho/\sigma^3$ tell us
something fundamental about the ROI or whether the Taylor-Navarro
result is, in some sense, accidental, remains to be seen.

\begin{figure}
\epsscale{1.0}
\plotone{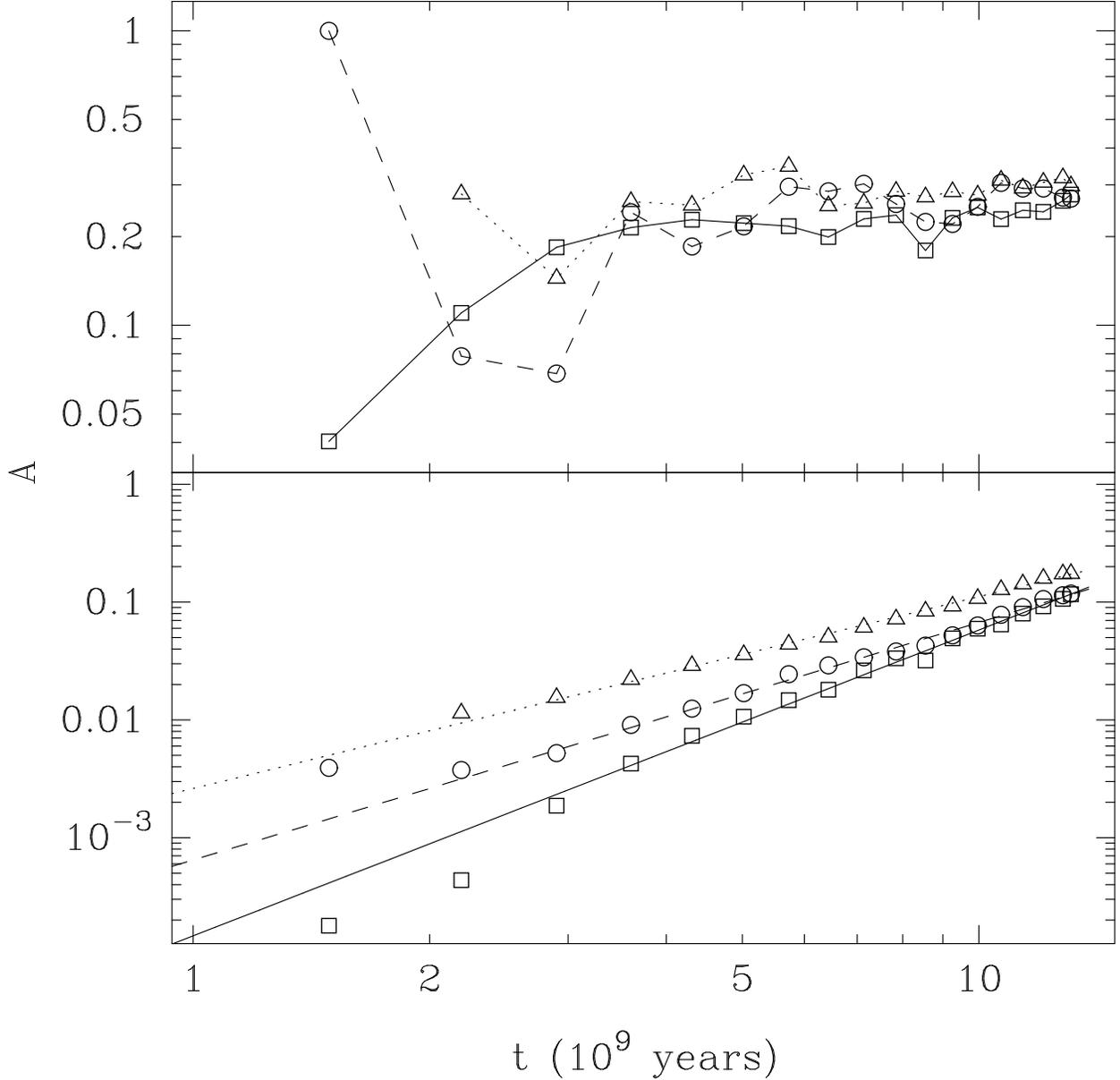}
\caption{Evolution of the bar strength.  The top panel shows $A$
calculated only within the virial radius $r_{200}$, while the bottom
panel computes $A$ over the entire system.  Squares and solid lines
denote Halo (b), circles and dashed lines denote Halo (d), and
triangles and dotted lines denote Halo (e).  The bottom panel also
displays least-square fits to the data, suggesting that $A \propto
t^\alpha$, with $\alpha = 2.60 \pm 0.03$ for the solid line, $\alpha =
2.02 \pm 0.01$ for the dashed line, and $\alpha = 1.63 \pm 0.02$ for
the dotted line.}
\label{fig:bar}
\end{figure}

The extent to which the ROI operates in a full cosmological simulation
(let alone the real Universe) may be regarded still as an open
question. However if the final density profiles have in fact to be
established very early, much doubt is removed.  Figure \ref{fig:bar}
suggests that the ROI drives isolated systems to a common final state
independent of perturbations, although at varying rates.  Less clear
is the relevance of coherent effects such as the ROI during major
mergers.  This and other questions will be left for future
investigations.
 

\acknowledgements{It is a pleasure to thank M. Duncan for useful
suggestions.  This work was supported, in part, by the Natural
Sciences and Engineering Research Council of Canada.}


\begin{deluxetable}{ccccccc}
\tablecaption{Collapsed halo parameters. \label{table:params}}
\tablehead{\colhead{Halo} & \colhead{Description} & \colhead{$M_{200}$ ($10^{12} \msun$)} & \colhead{$R_{200}$ (kpc)} & \colhead{$\lambda ^\prime$} & \colhead{$\beta$} }

\startdata
a & Smooth, radial & 2.56 & 354 & 0.0 & 1.0 \\
b & Smooth, unconstrained & 2.45 & 351 & $2.8 \times 10^{-4}$ & 0.81 \\
c & Perturbations, radial & 2.52 & 352 & $3.1 \times 10^{-5}$ & 0.99 \\
d & Perturbations, unconstrained & 2.38 & 346 & $3.8 \times 10^{-3}$ & 0.71 \\
e & Perturbations, external field & 2.39 & 347 & $2.6 \times 10^{-2}$ & 0.65 \\
\enddata
\label{tab:results}
\end{deluxetable}

\end{document}